\documentstyle[11pt,aaspp4]{article}
\def\gsim{\;\rlap{\lower 2.5pt
 \hbox{$\sim$}}\raise 1.5pt\hbox{$>$}\;}
\def\lsim{\;\rlap{\lower 2.5pt
   \hbox{$\sim$}}\raise 1.5pt\hbox{$<$}\;}
\newcommand\beq{\begin{equation}}
\newcommand\eeq{\end{equation}}

\begin{document}

\title{Microlensing of $\gamma$-Ray Burst Afterglows}
\author{Abraham Loeb and Rosalba Perna}
\medskip
\affil{Harvard-Smithsonian Center for Astrophysics, 60 Garden Street,
Cambridge, MA 02138}
%\altaffiltext{2}{email:aloeb@cfa.harvard.edu}

\begin{abstract}

The afterglow of a cosmological Gamma-Ray Burst (GRB) should appear on the
sky as a narrow emission ring of radius $\sim 3\times 10^{16}~{\rm
cm}~(t/{\rm day})^{5/8}$ which expands faster than light.  After a day, the
ring radius is comparable to the Einstein radius of a solar mass lens at a
cosmological distance.  Thus, microlensing by an intervening star can
modify significantly the lightcurve and polarization signal from a GRB
afterglow.  We show that the achromatic amplification signal of the
afterglow flux can be used to determine the impact parameter and expansion
rate of the source in units of the Einstein radius of the lens, and probe
the superluminal nature of the expansion.  If the synchrotron emission from
the afterglow photosphere originates from a set of coherent magnetic field
patches, microlensing would induce polarization variability due to the
transient magnification of the patches behind the lens. The microlensing
interpretation of the flux and polarization data can be confirmed by a
parallax experiment which would probe the amplification peak at different
times.  The fraction of microlensed afterglows can be used to calibrate the
density parameter of stellar-mass objects in the Universe.

\end{abstract}


%\centerline{submitted to {\it ApJ Letters}, August 1997}

\section{Introduction}

The recent discovery of delayed $X$-ray (van Paradijs et al. 1997),
optical (Bond 1997; Djorgovski et al. 1997; Mignoli et al. 1997), and
radio (Frail et al. 1997) emission, over hours to several months
following $\gamma$-ray bursts (GRB) established a new class of
variable sources in astronomy.  Of particular significance is the
detection of FeII and MgII absorption lines at a redshift of $z=0.835$
in the optical spectrum of GRB970508 (Metzger et al. 1997), which
confirmed the extragalactic origin of this burst. Since the source
redshift must be higher than the absorber redshift, its required
optical luminosity exceeds that of a supernova by several orders of
magnitude.  Thus, GRB afterglows might be detectable out to high
redshifts. One could then use the signatures of absorption in the
optical band (Metzger et al. 1997; Djorgovski et al.  1997),
scintillations in the radio regime (Goodman 1997), or gravitational
lensing (Gould 1992; Mao 1993) by intervening material along the
line-of-sight, to study the intrinsic properties of afterglow sources.

Afterglows are most naturally explained by models in which the bursts are
produced by relativistically expanding fireballs (Paczy\'nski \& Rhoads 1993;
Meszaros \& Rees 1997; Vietri 1996; Waxman 1997a,b; Wijers, Rees, \& Meszaros
1997; Vietri 1997; Sari 1997).  On encountering an external medium, the
relativistic shell which emitted the initial GRB
%(as a result of either internal or external shocks) 
decelerates and converts its bulk kinetic energy to synchrotron radiation,
giving rise to the afterglow.  The combined radio and optical data imply
that the fireball energy is $\sim 10^{51-52}$ erg.
%, and that the density of the medium into which the blastwave expands 
%is $\sim 1$ cm$^{-3}$. 
Due to relativistic beaming, the emission region seen by an external
observer, occupies an angle $\sim 1/\gamma$ relative to the center of the
explosion, where $\gamma$ is the Lorentz factor.  This region appears to
expand faster than the speed of light and occupies an angle of $\sim
0.1-10^2$ micro-arcseconds on the sky (or a physical size of $\sim
10^{15}$--$10^{18}~{\rm cm}$). Due to the smallness of this angular size,
it is difficult to resolve the afterglow source by terrestrial telescopes.
However, the lensing zone of a solar mass lens located at cosmological
distances occupies a micro-arcsecond on the sky (hence the term
``microlensing''), and thus offers the unique opportunity for resolving GRB
sources during their afterglow phase. Because of the superluminal expansion
of the source, any (non-relativistic) peculiar velocity of the lens
relative to the source can be ignored.  The amplification peak of a
microlensing event lasts for only $\la$ day, after which the net
amplification weakens as the source size grows larger than the Einstein
radius of the lens. The short duration of a microlensing event could
therefore provide a test for the high Lorentz factor of the afterglow
photosphere, which is predicted by all fireball models (for comparison, the
variations due to peculiar velocities in microlensing events of steady
sources take decades rather than days).

The rapid expansion and deceleration of the fireball causes a sharp decline
in its surface brightness as a function of time.  Since emission along the
line-of-sight to the source center suffers from the shortest geometric
time-delay, it occurs at larger radii and appears dimmer relative to
slightly off-axis emission. At any given time, the source is expected to
appear as a narrow ring of radius $R/\gamma$ and a width of order a tenth
of this radius (Waxman 1997c). The outer cut-off is set by the sharp
decline in relativistic beaming outside the ring. As the ring crosses a
lens, its magnification adds a sharp peak to the otherwise smooth light
curve of the afterglow.  The sharpness of the peak depends on the thickness
of the radiating gas layer behind the shock and on the shock deceleration
rate.  Microlensing could therefore provide important information about the
structure and dynamics of the afterglow photosphere.

The probability for stellar microlensing of a source at a redshift $z_s\sim
1$ is $\sim 0.1 \Omega_\star b^2$ (Press \& Gunn 1973; Gould 1995), where
$\Omega_\star$ is the mean density of stellar-mass objects in the Universe
in units of the critical density, and $b$ is the impact parameter of the
source relative to the lens in units of the Einstein radius.  The known
population of luminous stars amounts to $\Omega_\star\sim 5\times 10^{-3}$
(Woods \& Loeb 1997), and implies that most cosmological sources are
separated from stellar lenses by $b\sim 40$.  The typical impact parameter
is smaller by an order of magnitude if the dark matter is made of Massive
Compact Halo Objects (MACHOs) as Galactic microlensing searches suggest
(Alcock et al.  1996).

In this paper we examine the question whether a stellar mass lens can
resolve the predicted properties of afterglow photospheres.  For
concreteness, we derive numerical results for the fireball emission model
of Waxman (1997b,c).  Since the afterglow occurs long after the explosive
energy release, its properties are not sensitive to the spatial or temporal
details of the point explosion that triggered the GRB.  However, our adopted
emission model is by no means a unique interpretation of the existing
afterglow data (see, e.g. Vietri 1997 or Paczy\'nski 1997); in fact, a
future detection of a microlensing signal could serve to discriminate among
competing afterglow models.

In \S 2 we describe our model for GRB afterglows and characterize both the
intensity and polarization signals that would result from a microlensing
event. The numerical results and their implications are discussed in \S 3.
Finally, \S 4 summarizes the main conclusions from this work.

\section{Source Model and Microlensing Signatures}

\subsection{Source Model}

To illustrate the effects of microlensing on GRB afterglows we need to
specify the evolution of the source size and spectral intensity with time.
We adopt the scaling laws for the expansion and emission of a relativistic
fireball which decelerates in a uniform ambient medium (Waxman 1997b).

In the fireball model, a compact ($\sim 10^{6-7}$ cm) source, releases an
energy of $E\sim 10^{52}$ ergs over $T\la 10^2$ seconds with a negligible
baryonic contamination ($\la 10^{-5}M_\odot$). The high energy density at
the source results in an optically thick pair plasma that expands and
accelerates to relativistic velocities. After an initial acceleration
phase, the thermal energy is converted to kinetic energy of the protons. A
cold shell of thickness $cT$ is formed and continues to expand. Internal
shell collisions as a result of unsteady source activity could convert part
of kinetic energy into radiation and yield the primary GRB emission via
synchrotron emission and inverse-Compton scattering (Paczy\'nski \& Xu
1993; Meszaros \& Rees 1994; Sari \& Piran 1997).  As the cold shell
expands, it impacts on the surrounding medium and drives a
relativistic shock also in it; this shock continuously heats fresh gas and
accelerates relativistic electrons which produce via synchrotron emission
the delayed radiation observed on time scales of hours to months.
Following Waxman (1997b), the radius of the shock at observed time $t$ is
given by
\begin{equation}
R(t)\approx 8.7\times 10^{16} E_{52}^{1/4} n_1^{-1/4} t_{\rm
hr}^{1/4}\;{\rm cm}
\label{eq:r}
\end{equation}
while its Lorentz factor is
\begin{equation}
\gamma(t)=\sqrt{\frac{R(t)}{2ct}}\approx
21E_{52}^{1/8}n_1^{-1/8}t_{\rm hr}^{-3/8}\;.
\label{eq:gamma}
\end{equation}
Here $E_{52}$ is the fireball energy in units of $10^{52}$ ergs, $n_1$ is
the ambient gas density in cm$^{-3}$ and $t_{\rm hr}$ is the
observed time in hours.  As mentioned in \S 1, 
most of the emission is seen from a narrow ring of radius
\begin{equation}
\rho_s(t)=\frac{R(t)}{\gamma(t)}\approx 4.1\times 
10^{15} E_{52}^{1/8} n_1^{-1/8} 
t_{\rm hr}^{5/8}\;{\rm cm} \;.
\label{eq:rs}
\end{equation}
The width of the ring is a fraction $W\sim 10\%$ of its radius $\rho_s$ if
the thickness of the radiating layer behind the shock is determined by the
shock hydrodynamics in a self-similar expansion (Waxman 1997c).  A thicker
radiating layer (e.g. due to a large gyroradius of the radiating electrons)
would result in a wider ring. In \S 3, we will show numerical results for
different choices of $W$.
These expressions are valid also for a jet geometry as long as the opening
angle of the jet is $\ga 1/\gamma$.

The $X$-ray, optical and radio emission following the $\gamma$-ray burst
can be modelled as synchrotron emission from a power-law population of
electrons within the heated shell behind the expanding shock. Under the
assumption that the magnetic field energy density in the shell rest frame
is a fraction $\xi_B$ of the equipartition value, and that the power-law
electrons carry a fraction $\xi_e$ of the dissipated energy, the observed
frequency at which the synchrotron spectral intensity of the electrons
peaks is
\begin{equation}
\nu_m(t)=5.9\times 10^{15}\left(\frac{1+z_s}{2}\right)^{1/2}
\left({\xi_e\over 0.1}\right)^2 
\left({\xi_B\over 0.1}\right)^{1/2} 
E_{52}^{1/2}t_{\rm hr}^{-3/2}\;{\rm Hz}\;,
\label{eq:num}
\end{equation}
where $z_s$ is the cosmological redshift of the source. The observed
intensity at $\nu_m$ is
\begin{equation}
F_{\nu_m}=1.0\left(\frac{1+z_s}{2}\right)^{-1}\left[\frac{1-1/\sqrt{2}}
{1-1/\sqrt{1+z_s}}\right]^2 n_1^{1/2}\left({\xi_B\over 0.1}\right)^{1/2}
E_{52}\;{\rm mJy}\;.
\label{eq:fnum}
\end{equation}
If the distribution of electron Lorentz factors follows a power-law,
$dN_e/d\gamma_e
\propto \gamma_e^{-p}$, with a low energy cut-off set by $\xi_e$, 
then the observed intensity as a function of frequency, $\nu$, obeys,
\begin{equation}
F^0_\nu(t) = F_{\nu_m}\left[\nu/\nu_m(t)\right]^{-\alpha}\;,
\label{eq:fnu}
\end{equation}
where $\nu_m$ is the emission frequency of the electrons at the low-energy
cut-off. The variation in $\nu_m$ across the finite width of the ring can
be ignored for $W\ll 1$.  The typical parameter values which are required
to fit the afterglow data are $\xi_e\sim0.1$, $\xi_B\sim 0.1$, and $p\sim
2$, so that $\alpha\sim 0.5$ for $\nu >
\nu_m$ and $\alpha=-1/3$ for $\nu < \nu_m$.

\subsection{Flux Amplification Due to Microlensing}

We now consider a point lens of mass $M$ and redshift $z_l$ which happens
to be located near the line-of-sight to an expanding fireball.  We denote
by $\eta$ the impact parameter of the source center relative 
to the observer-lens axis.
% and by $R_s(t)$ the radius of the source at time $t$. 
For simplicity we assume that the source has a uniform surface brightness
in a ring of radius $\rho_s(t)$ [given by equation~(\ref{eq:rs})] and a
width $W\rho_s(t)$.  The flux seen by the observer is
\begin{equation}
F_\nu^{\rm lens}[t,R_s(t),W,b]=F^0_\nu(t)\;\mu[R_s(t),W,b]\;,
\label{eq:flens}
\end{equation}
where $R_s\equiv \rho_s/r_{\rm E}$, and $r_{\rm E}$ is the Einstein radius
of the lens projected on the source plane, $r_{\rm
E}=\sqrt{(4GM/c^2)(D_{\rm s} D_{\rm ls}/D_{\rm l})}$, with $D_{\rm
l}$,$D_{\rm s}$ and $D_{\rm ls}$ being the angular diameter distance to the
lens, to the source, and from the lens to the source, respectively. These
distances all depend on the cosmological parameters.  In this paper we
assume $\Omega=1$, $\Lambda=0$, and $H_0=50~{\rm km~s^{-1}~Mpc^{-1}}$. The
magnification factor for a normalized lens-source separation
$b\equiv\eta/r_{\rm E}$ is,
\begin{equation}
\mu(R_s,W,b)={{\Psi}[R_s,b]-(1-W)^2{\Psi}[(1-W)R_s,b]\over 1 - (1-W)^2},
\label{eq:mu}
\end{equation}
where ${\Psi}(R_s,b)$ is the magnification for a uniform disk of radius
$R_s$ (Schneider, Falco, \& Ehlers 1992),
\begin{equation}
{\Psi}[R_s,b]= \frac{2}{\pi R_s^2}\left[\int_{|b-R_s|}^{b+R_s} dr
\frac{r^2+2}{\sqrt{r^2+4}}\arccos\frac{b^2+r^2-R_s^2}{2rb} +
H(R_s-b)\frac{\pi}{2}(R_s-b)\sqrt{(R_s-b)^2+4}\right]\;.
\label{eq:circ}
\end{equation}
Here $H(x)$ is the Heaviside step function.  The integral in equation
(\ref{eq:circ}) can be expressed more explicitly as a sum of elliptic
integrals (Witt \& Mao 1994). 
%For a narrow ring with $W\ll1$, the
%magnification $\mu\approx \Psi(R_s) + {1\over 2}R_s
%\left(d\Psi/dR_s\right)\vert_{R_s}$. 
Other analytic results exist for more general surface brightness
distributions (Heyrovsk\'y \& Loeb 1997).

\subsection{Polarization Variability Due to Microlensing}

If the afterglow photosphere contains a finite set of discrete patches,
each having a coherent magnetic field distribution, then the emergent
synchrotron radiation will be polarized.  For a power-law distribution of
electron energies with an index $p$, the degree of linear polarization in
each coherent patch is given by (Rybicki \& Lightman 1979)
\begin{equation}
\Pi =\frac{p+1}{p+7/3}\;.
\label{eq:pi}
\end{equation}
For the inferred value of $p\sim 2$ (Waxman 1997b), $\Pi\sim 0.7$.  A
microlens capable of resolving the source, could then provide useful
information about its magnetic field structure.

To illustrate the effect of microlensing on polarization we adopt a toy
model in which the emission ring is divided into a set of independent
segments, each having a coherent distribution of magnetic field lines.  The
polarization in each segment is then modelled as a traceless symmetric
$2\times2$ tensor with a random orientation angle, and a contraction
$(P_{\alpha\beta}P^{\alpha\beta})^{1/2}=\Pi$, given by equation~(\ref{eq:pi}).
To simplify the computation, we subdivide the emission ring into $N$
segments of equal area and nearly square shape.
To each segment we assign a randomly oriented linear polarization.  For the
sake of concreteness, we assume that the number of segments and the
orientation of their polarization stays constant during the lensing event.
This assumption is reasonable since the effect of lensing peaks during the
short period of time when the ring crosses the lens (which is smaller than
the expansion time by a factor $W\ll1$).

The net observed polarization is then given by
\begin{equation}
\langle {\vec{\vec{P}}}\rangle=
\frac{\sum_{i=1}^N {\vec{\vec{P}}}_{i}A_{i}}{\sum_{i=1}^N A_{i}}\;,
\label{eq:pol}
\end{equation}
where $A_i$ and ${\vec{\vec{P_i}}}$ are the area 
and polarization tensor of the $i$-th segment, and
\begin{equation}
{\vec{\vec{P}}}_{i}={\Pi\over \sqrt{2}}
\left(\matrix{\cos2\phi_{i}&\sin2\phi_{i}\cr\sin2\phi_{i}
&-\cos2\phi_{i}}\right),
\label{eq:matpol}
\end{equation}
with a random orientation angle, $0\leq\phi_{i}<2\pi$.
 
We first consider the case where there is no lensing, in which
$A_{i}=A_0=const$ for $i=1,...N$. The two components of the net
polarization are then given by
\begin{eqnarray}
\langle P\rangle_{xx}=-\langle P\rangle_{yy} &=&\frac{\sum_{i=1}^N \Pi
\cos{2\phi}_{i}A_{0}}{{\sqrt{2}}N A_0}
= \frac{\Pi}{\sqrt{2}N}\sum_{i=1}^N \cos{2\phi}_{i} \nonumber\\
\langle P\rangle_{xy}=\langle P\rangle_{yx}&=&
\frac{\sum_{i=1}^N \Pi\sin{2\phi}_{i}A_{0}}{{\sqrt{2}}N A_0}
= \frac{\Pi}{\sqrt{2}N}\sum_{i=1}^N \sin{2\phi}_{i}\;.
\label{eq:pol0}
\end{eqnarray}
Clearly, the resulting polarization $\langle P\rangle
=\sqrt{2(\langle P\rangle_{xx}^2+\langle P\rangle^2_{xy})}$
approaches zero for large $N$ and is time independent.

Let us now consider the situation where a lens is located at a projected
position $(x_l,y_l)$ with respect to the center of the source, so that
$b=(x_l^2+ y_l^2)^{1/2}$.  The observed polarization is still given by
equation~(\ref{eq:pol}) but due to the stretching caused by lensing, the
areas $\{A_{i}\}_{i=1}^{N}$ of the various segments are no longer equal,
\begin{equation}
A_{i}(t)=\int_{i}\int \zeta d\zeta d\theta
\frac{d^2+2}{d\sqrt{d^2+4}},
\label{eq:aij}
\end{equation}
where $(\zeta,\theta)$ are polar coordinates centered on the source, and
the integrand is the point-source amplification factor at an impact
parameter $d\equiv [{(\zeta\cos\theta -x_l)^2+(\zeta\sin\theta
-y_l)^2}]^{1/2}$.  The integral is taken over the unlensed area of the
segments.  Because the size and position of the various segments relative
to the lens change with time, the observed polarization will vary during a
microlensing event.

\section{Numerical Results}

\subsection{Flux Amplification}

The solid lines in Figure 1 show the unlensed $10^{14}$Hz flux of an
afterglow according to equation~(\ref{eq:fnu}) and the parameter choices
mentioned below that equation.  The broken lines show the effect of
microlensing on the observed flux, according to equation~(\ref{eq:flens}),
for different choices of the ring's fractional width $W$ and impact
parameter $b$.

The qualitative features of the microlensing signature on the afterglow
lightcurve are as follows:

\begin{itemize}
\item[(a)] All wavelengths show the same amplification profile
as a function of time\footnote{The variation of the relativistic Doppler
effect across the ring might result in a slight chromaticity of the lensing
signal, but we ignore it here.}. While the amplification peak occurs on the
rising side of the lightcurve in the radio, it appears on its declining
side in X-rays, and might show on both sides of the break in the optical
[see Eq.~(\ref{eq:num}) for the timing of the peak at a given frequency].
The larger $b$ is, the easier it becomes to detect the amplification signal
at longer wavelengths.  For example, the optimal frequencies for detecting
the signals shown in Figure 1c and 1d are $10^3$ GHz (sub-mm) and $10^2$
GHz (radio), respectively.  The achromaticity of the amplification peak can
be used to separate the lensing signal from noise due to intrinsic
variability or interstellar scintillations.  Detailed observations of
future afterglows are necessary in order to assess the characteristic level
of intrinsic variability.

\item[(b)] At early times, the temporal profiles of the lensed and unlensed
fluxes have the same shape but different amplitudes.  During this period,
the source can still be regarded as pointlike and the offset between the
lensed and unlensed curves is set by the point source magnification
factor at a constant $b$. The unknown value of $b$ could therefore be
inferred from this asymptotic offset in amplitude between the lensed and
unlensed regimes.

\item[(c)]
The maximum amplification occurs at the time $t_\star$ when the ring crosses
the lens, namely when $R_s\sim b$. The otherwise unknown source size $R_s$
can therefore be inferred at the time $t_\star$.  By taking the ratio between
the ring size and the period $t_\star$, one finds the mean velocity of the
expanding ring during that time interval in units of the Einstein radius,
$r_E$.  Given a probability distribution for $r_E$ (based on a reasonable
mass and redshift distribution for the lenses), one could then test the
hypothesis of superluminal expansion.

\item[(d)]
Analysis of the shape of the lightcurve after the peak can provide more
detailed information about the fractional width of the ring $W$ and the
temporal history of $R_s(t)$. The smaller $W$ is, the higher and narrower
the amplification peak gets.  When $R_s\sim b$, the value of the
magnification $\mu(R_s,W,b)$ becomes highly sensitive to the source size
$R_s$.  By monitoring the lensed flux as a function of time, one could
infer the magnification $\mu_{\rm obs}(t) =F_{\rm obs}(t)/F_0(t)$, where
$F_0(t)$ is found from the power-law extrapolation of the observed
$F_{\rm obs}(t)$ after the end of the microlensing event, to earlier times.
Based on the magnification history $\mu_{\rm obs}(t)$ one could infer the
time evolution of $R_s(t)$ from the constraint $\mu(R_s,W,b)=\mu_{\rm
obs}(t)$, where $\mu(R_s,W,b)$ is given by equation~(\ref{eq:mu}) and $b$
is inferred based on point (b) above.

\end{itemize}

The quantitative interpretation of the lensing signatures suffers from an
ambiguity about the physical size of the Einstein radius of the lens.  This
ambiguity can be removed through a parallax experiment, in which two (or
more) telescopes, separated across the solar system, observe the same
microlensing event with different values of $b$ (Grieger, Kayser, \&
Refsdal 1986; Gould 1994).  Since variability is induced by the
superluminal expansion of the source (rather than by the motion of the
lens, as is usually the case in microlensing events of steady sources), the
two telescopes would simply observe different lightcurves with different
values of $b$. Based on their known separation and their inferred $b$
values [see point (b)] one could then measure the physical size of the
Einstein radius, $r_{\rm E}$. The shape of the different peaks measured by
the two telescopes can be used to test for self-similarity in the shock
structure and dynamics.

\subsection{Microlensed Polarization}

The different lines in Figure 2 show the deviation from the steady
polarization signal that is predicted by equation~(\ref{eq:pol0}), due to
microlensing [Eqs. (\ref{eq:pol}) and (\ref{eq:aij})].  The different
panels show several random realizations for various choices of the
lens-source separation $b$, and the number of ring segments $N$.  We
consider two values of $N$, one in which the ring is composed of a single
radial strip composed of nearly square segments ($N=63$), and a second in
which it is divided into two such strips ($N=250$).  The particular value
that the polarization obtains at any given time $t$ depends on the specific
set of random orientation angles $\{\phi_{i}\}_{i=1}^{N}$ that were
assigned to the segments in each realization, and so the fluctuations
induced by lensing should be analyzed on a statistical basis only.

The main qualitative characteristics of the lensing signal are:

\begin{itemize}

\item[(a)] The polarization changes around $t=t_\star$ in coincidence 
with the flux amplification peak. At that time, the polarization fluctuates
because as the ring expands, different segments approach the lens (and
hence the point of maximum amplification) at different times. At any given
time, the segment which crosses the lens obtains the largest area in the
image plane and provides the largest contribution to the overall
polarization. The fluctuation rate increases as the area of each individual
segment gets smaller (or as $N$ gets larger), because smaller segments
sweep faster across the lens.

\item[(b)]
If the ring is narrower than the Einstein diameter at lens crossing (i.e.
$Wb\la 1$), then the typical fluctuation amplitude, $\delta\equiv (\langle
P\rangle/\langle P_0\rangle) -1$, is roughly independent of $N$ (see top
panels of Fig. 2).  In this case, the ring is sliced into a fixed number of
``effective'' segments, each having a length of order the Einstein
diameter, so that $N_{\rm eff}\sim (2\pi\rho_s)/(2r_{\rm E})\sim \pi b$,
and $\delta \sim N_{\rm eff}^{-1/2}$.  

\item[(c)]
The fluctuation amplitude decreases with increasing $b$, because in this
limit the highly-magnified zone behind the lens amounts to a smaller
fraction of the entire ring area.

%\item[(d]) The shoulders of the polarization extrema are broader 
%the smaller $b$ is, because in that limit the lensing effect remains
%significant even when the ring does not perfectly overlap with the lens.

%\item[(d)]
%For a particular choice of $N$ and $b$, the duration of a typical
%fluctuation increases with time. The expansion of the source causes a
%steady increase in the size of its segments and a corresponding lengthening
%of the period it takes a given patch to cross the Einstein radius of the
%lens.  
%This result is quantified in Figure 3, where we plot the expected
%number of extrema (either maxima or minima) in the polarization signal as a
%function of time.  All curves are characterized by an initial rise, up to
%the time when $R_s\sim b$.  The subsequent decline is due to the increase
%in segment size, and the corresponding lengthening in the amount of time
%required for a single segment to sweep across the lens.

\end{itemize}

\section{Conclusions}

We have shown that microlensing by stars can be used to study the size,
superluminal expansion rate, and granularity of the photospheres of GRB
afterglows.  The light curves shown in Figure 1 can be used to extract the
source impact parameter $b$ relative to the lens (based on the
normalization offset between the pre- and post-lensing curves), the
fractional width of the emission ring (from the height and width of the
amplification peak), and the source expansion rate and size in units of the
Einstein radius of the lens.  The source size can be measured explicitly
through a parallax experiment which would obtain two (or more) light curves
that sample the achromatic amplification peak at different times (cf. Fig.
1).  Such an experiment could serve as the definitive tool for
discriminating between a microlensing event and intrinsic variability of
the afterglow source.

By monitoring the variability of the polarization with time during a
microlensing event, it is also possible to estimate the number of coherent
magnetic field patches on the afterglow photosphere (Fig. 2).  
%The polarization variability shown in Figures 1 and 2 for $b\sim 10$ should
%appear in at least one out of twenty GRB afterglow, based on the known
%population of luminous stars in the Universe.  

If the cosmological density parameter of stellar mass MACHOs is
$\Omega_\star$, then most afterglow events will acquire an impact parameter
$b\la 10 (\Omega_\star/0.1)^{-1/2}$ from their nearest lens.  Multi-band
photometry with an accuracy of $\sim 0.03$ mag, could then detect the flux
amplification signal shown in Figures 1a-c and test for its achromaticity,
or else place interesting upper limits on $\Omega_\star$, based on a
relatively small sample of frequently-monitored afterglows.  The $\sim 1\%$
amplification signal shown in Figure 1d for $b=10$ would appear in 5--100\%
of all afterglows after 2--3 months, at the time when the peak flux of
$\sim$ mJy is reached in the radio, at $\sim 10^2$GHz.  A future X-ray
satellite which would locate afterglows to within an arcminute (like
BeppoSAX does) for all GRBs detected by BATSE, might identify hundreds of
afterglows per year, and could provide a rich sample for such microlensing
studies.

Although our results were limited to isolated point lenses, their
qualitative features should be common to lens systems with more complicated
caustic structure, such as binary stars or galactic cores.

\acknowledgements

We thank Eli Waxman for valuable discussions and for communicating results
from his work prior to publication, and David Heyrovsk\'y for useful
comments on the manuscript.  This work was supported in part by the NASA
ATP grant NAG5-3085 and the Harvard Milton fund.

%\newcounter{figmain}
%\newcounter{figsub}[figmain]
%\renewcommand{\thefigure}{\arabic{figmain}.\alph{figsub}}
%\refstepcounter{figmain}
%\refstepcounter{figmain}
%\refstepcounter{figmain}  

%\refstepcounter{figsub}
\begin{figure}[t]
\centerline{\epsfysize=5.7in\epsffile{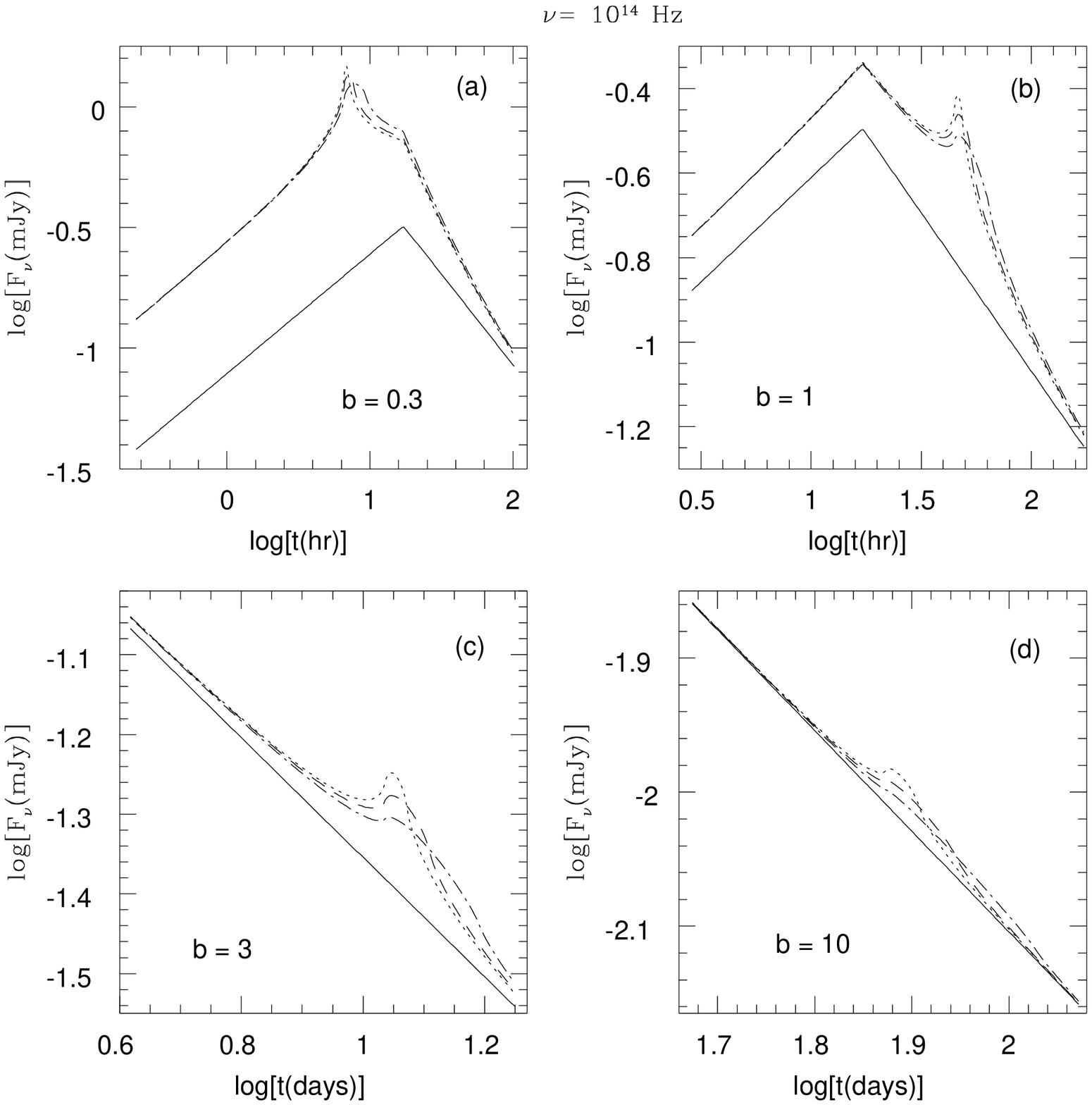}}
\caption{The unlensed (solid line) and lensed (broken lines)
flux from a GRB afterglow with $E_{52}=1$ and $n_1=1$ at a frequency
$\nu=10^{14}~{\rm Hz}$.  The different broken lines correspond to different
fractional widths of the emission ring, $W=5\%$ (short-dashed, highest
peak), $10\%$ (long-dashed, middle peak) and $20\%$ (dot-dashed, lowest
peak).  The lens mass is $M=1M_\odot$ and its redshift is $z_l=0.5$.  The
source redshift is $z_s=2$. The likelihood for the events shown is $\sim
(10$--$30)\% (\Omega_\star/0.1)(b/3)^2$ (see Fig. 1 in Gould 1995). }
\label{fig:1}
\end{figure}

%\refstepcounter{figsub}
%\begin{figure}[t]
%\centerline{\epsfysize=5.7in\epsffile{/h4/perna/GammaRays/fig1b.ps}}
%\caption{Same as Figure 1a, but with $M=M_\odot$.}
%\label{fig:1b}
%\end{figure}

%\refstepcounter{figsub}
%\begin{figure}[t]
%\centerline{\epsfysize=5.7in\epsffile{fig1c.ps}}
%\caption{Same as Figure 1a, but with $M=10 M_\odot$.}
%\label{fig:1c}
%\end{figure}

%\refstepcounter{figmain}
%\refstepcounter{figsub}

\begin{figure}[t]
\centerline{\epsfysize=5.7in\epsffile{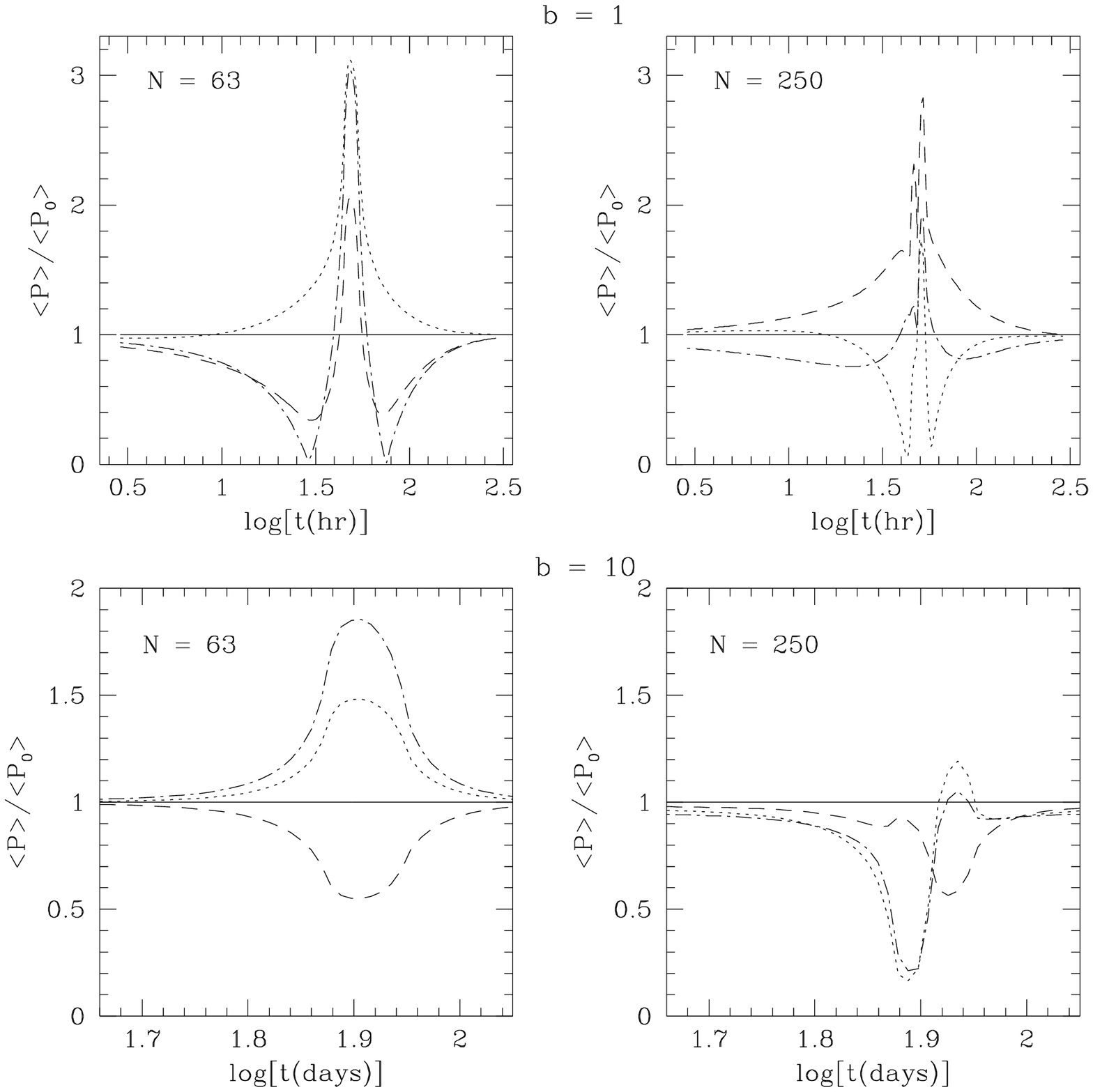}}
\caption{The lensed polarization signal, $\langle P\rangle$, normalized by the
(constant) unlensed value $\langle P_0\rangle$.  The different lines show
three random realizations of the time-varying polarization that would be
observed during a microlensing event if the emission ring is composed of
$N$ nearly-square segments which produce a polarization of equal amplitude
but random orientation. Results are shown for different values of $N$ and
the source-lens separation $b$.  The unlensed polarization is $\langle
P_0\rangle\approx 0.09$ and $0.04$ for $N=63$ and $250$, respectively.
Parameters are the same as in Figure 1 with $W=10\%$.}
\label{fig:2}
\end{figure}
%\refstepcounter{figsub}
%\begin{figure}[t]
%\centerline{\epsfysize=5.7in\epsffile{fig2b.ps}}
%\caption{Same as in Fig.2a, but for a given $N$ and varying $b$.}
%\label{fig:2b}
%\end{figure}

%\renewcommand{\thefigure}{\arabic{figmain}}
%\setcounter{figmain}{3}
%\begin{figure}[t]
%\centerline{\epsfysize=5.7in\epsffile{/h4/perna/GammaRays/fig3.ps}}
%\caption{Average number (over 20 realizations) of extrema 
%in the temporal fluctuations of the polarization signal from a microlensed
%afterglow.  Parameters are the same as in Figure 1, and $W=10\%$. }
%\label{fig:3}
%\end{figure}

\end{document}